\documentclass[10pt,conference]{IEEEtran}

\usepackage{cite}
\usepackage{amsmath,amsthm,amsfonts,amssymb,amsbsy,esint,cancel}

\usepackage{graphicx}
\usepackage{xcolor}

\usepackage{textcomp,gensymb,bbm,bm,dsfont,upgreek}

\usepackage[caption=false,font=footnotesize]{subfig}

\usepackage{url} 
\usepackage[bookmarks=false,hidelinks]{hyperref} 

\usepackage[normalem]{ulem} 

\usepackage{tikz}
\tikzset{every picture/.style={font issue=\footnotesize},
	font issue/.style={execute at begin picture={#1\selectfont}}
}
\usepackage{pgfplots}
\pgfplotsset{compat=newest}
\usetikzlibrary{backgrounds,plotmarks}
\usetikzlibrary{positioning,calc} 
\usetikzlibrary{arrows,arrows.meta}
\usepackage{ulem} 

\usepackage{mathtools}

\usepackage[capitalize]{cleveref}
	\Crefname{figure}{Fig.}{Fig.}
	\Crefname{section}{Sec.}{Sec.}
	\Crefname{subsection}{Sec.}{Sec.}
	\Crefname{prop}{Proposition}{Proposition}
	\Crefname{lemma}{Lemma}{Lemma}
	\Crefname{equation}{}{}
	\Crefname{footnote}{Footnote}{Footnote}

  \newcommand\f[2]{\frac{#1}{#2}} 
  \newcommand\re{\mathrm{Re}} 
  \newcommand\im{\mathrm{Im}} 
\renewcommand\Re{\re} 
\renewcommand\Im{\im} 

\DeclareMathOperator*\supp{supp}

\renewcommand\H{^\mathrm{\normalfont{H}}}        

\newcommand\bbN{\mathbb{N}} 
\newcommand\bbZ{\mathbb{Z}} 
\newcommand\bbR{\mathbb{R}} 
\newcommand\bbC{\mathbb{C}} 

\newcommand\SymbolOfEV{\mathbb{E}} 
\newcommand\EVVar[2]{\SymbolOfEV_{#1}[\hspace{.2mm}#2\hspace{.2mm}]}
\newcommand\EV[1]{\EVVar{}{#1}}

\newcommand\iid{\overset{\text{i.i.d.}}{\sim}}

\newcommand\calU{\mathcal{U}}   
\newcommand\calN{\mathcal{N}}   
\newcommand\calCN{\mathcal{CN}} 
  
\newcommand\unit[1]{\,\mathrm{#1}}	
\newcommand\dB{\unit{dB}}

\newcommand\SymbolOfTx{\hspace{.2mm}\mathrm{T}}
\newcommand\SymbolOfRx{\hspace{.2mm}\mathrm{R}}
\newcommand\Txs[1]{_{\SymbolOfTx#1}} 
\newcommand\Rxs[1]{_{\SymbolOfRx#1}} 
\newcommand\RxTxs[1]{_{\hspace{.2mm}\mathrm{RT}#1}}
\newcommand\Tx{\Txs{}} 
\newcommand\Rx{\Rxs{}} 
\newcommand\RxTx{\RxTxs{}} 
\newcommand\SNR{\mathrm{SNR}}

\newcommand\iTx{i\Tx}
\newcommand\ChannelFunc{g} 
\newcommand\Radius{i}
\newcommand\Disk{\mathcal{D}_i}
\newcommand\HalfPlane{\mathcal{H}_z}
\newcommand\DiskRadius{\Radius_1}

\newcommand\SymbolAlphabet{\mathcal{I}}
\newcommand\SymbolAlphabetZL{\mathcal{Z}}
\newcommand\Rate{I(y;i)} 
\newcommand\RateMax{C}
\newcommand\calC{\mathcal{C}} 

\newcommand\CircProb{q}
\newcommand\SymbProb{p}

\renewcommand\L{_\mathrm{L}}     
\newcommand\Res{_\mathrm{res}}   
\newcommand\Ind{^\mathrm{ind}}   
\newcommand\Amb{^\mathrm{ext}}   
\newcommand\N{_\mathrm{N}}

\begin{document}

\newcommand\OurPaperTitle{%
The Channel Capacity of General Complex-Valued Load Modulation for Backscatter Communication%
}
\title{\OurPaperTitle}
\author{
\IEEEauthorblockN{Gregor Dumphart, Johannes Sager, and Armin Wittneben} 
\IEEEauthorblockA{\textit{Wireless Communications Group, D-ITET, ETH Zurich, Switzerland}\\
Email: dumphart@nari.ee.ethz.ch, sagerj@student.ethz.ch, wittneben@nari.ee.ethz.ch}}

\maketitle

\begin{abstract}
%
This paper studies achievable information rates of backscatter communication systems where the tag performs load modulation with a freely adaptable passive termination. We find that the complex phasor of the tag current is constrained to a disk and that the capacity problem can therefore be described with existing results on peak-power-limited quadrature channels. This allows us to state the channel capacity and the capacity-achieving distribution of the load impedance, which is described by non-concentric circles in the right half-plane. For the low-SNR case (SNR \(<\) 4.8 dB) we find that channel capacity is achieved by a purely reactive load with Cauchy-distributed reactance. The exposition is based on a system model that abstracts all relevant classes of backscatter communication systems, including RFID. To address practicality, we construct a symbol alphabet that allows for a near-capacity information rate of more than 6 bit per load-switching period at reasonably high SNR. We also find that the rate hardly decreases when typical value-range constraints are imposed on the load impedance.
\end{abstract}

\newcommand\OurKeywords{%
load modulation, backscatter communication, RFID, ambient backscatter, channel capacity, achievable rate, low energy communication, passive communication}


\section{Introduction}
\label{sec:intro}
Backscatter communication (BC) via load modulation allows simple passive tags to communicate with essentially zero transmit power and no transmit amplifier. This is achieved by modulating the termination load of the tag antenna in order to affect the reflection of an incident field (possibly an ambient field). This technique found widespread use in radio-frequency identification (RFID) and smart cards \cite{Finkenzeller2015} and is a promising approach to ultra-low-energy communication in the Internet of Things (IoT) \cite{HuynhCST2018}. The high data rate requirements of many IoT applications have recently prompted interest in backscatter modulation beyond binary \cite[Tab.~III]{HuynhCST2018}, e.g. 16-QAM \cite{KimionisNAT2021} or QPSK \cite{WangMercierISSCC2020}, together with error-correcting codes \cite{HuynhCST2018}. 

From the perspective of communication theory, it is natural to ask for the channel capacity of a BC link, i.e. the maximum achievable information rate.
The existing research literature contains only a few related investigations. For example \cite{ZhaoACCESS2018} addresses the calculation of the channel capacity of binary load modulation in ambient backscatter communication (ABC) for various cases of the ambient signal modulation.
The work in \cite{KimSPAWC2017} concerns the maximization of ABC network capacity in terms of redundancy and reflection coefficient (for BPSK, QPSK, and 16-QAM alphabets) in a WiFi setting with OFDM. The focus of \cite{FuschiniAPL2008} is on the effect of the propagation environment on the Euclidean symbol distances and the resulting bit error rate with PSK and ASK for RFID load modulation.
The literature lacks a complete description of the channel capacity and the capacity-achieving transmit scheme of BC load modulation, which would provide a crucial guideline for the design of practical systems with near-optimal rates \cite{Tse2005}.

This paper describes for the first time the channel capacity of load modulation in the general case of a  freely adaptable passive load. In this case, the load impedance can take on any complex value with non-negative real part for the duration of every symbol period. 
This is a generalization of specific modulation schemes such as QPSK, where the load takes values from a finite alphabet. The results and insights promise useful implications for practical BC systems.

This paper contains the following specific contributions:
\begin{itemize}
\item We develop a signal and noise model that abstracts all major classes of load-modulated single-tag BC links.
\item Based thereon, we study the physical constraints on the tag-side transmit signal, arising from the passive nature of the tag. We find that the transmit current phasor $i \in \bbC$ must lie in a certain disk in the right half-plane. 
\item We discover that this disk constraint allows to solve the channel capacity problem at hand with existing theory on peak-power-constrained quadrature AWGN channels.
\item The capacity result is stated and discussed. We identify special cases in which the result even applies to ABC.
\item The capacity-achieving distribution of the transmit current and of the load impedance are characterized in detail. For the low-SNR case we show that a purely reactive load with Cauchy-distributed reactance achieves capacity.
\item We construct a finite symbol alphabet that approximates the capacity-achieving distribution. It yields near-capacity data rates, even if several symbols are unrealizable due to implementation constraints on the load.
\end{itemize}

This paper does not address the tag power consumption or aspects of the energy harvesting circuit. Specific channel models and multi-user interference are also out of scope.

\subsubsection*{Paper Structure}
\Cref{sec:model} describes the employed system model and \Cref{sec:constraint} the special transmit-side constraints. \Cref{sec:capacity} states the channel capacity, the associated distributions, and a familiar upper bound. \Cref{sec:practical} addresses practical modulation aspects and \Cref{sec:summary} concludes the paper.

\subsubsection*{Notation}
For a random variable $x$, the probability density function (PDF) is denoted as $f_x(x)$. For simplicity, we do not use distinct random variable notation.

\section{System Model}
\label{sec:model}
\begin{figure}[!ht]
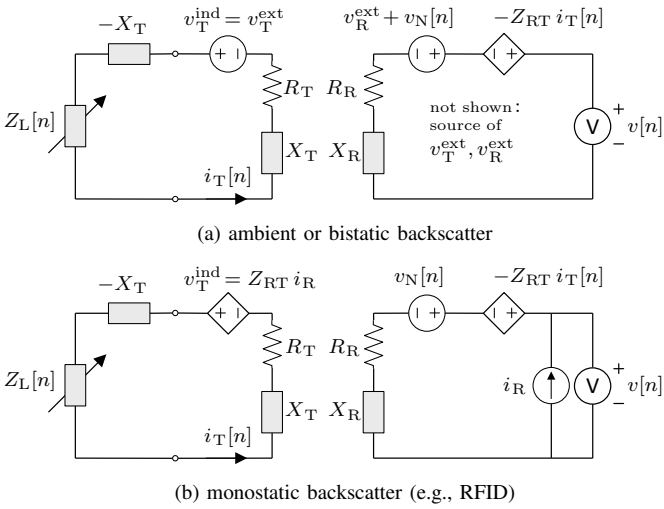

\centering
\subfloat[ambient or bistatic backscatter]{\centering
\resizebox{.9\columnwidth}{!}{\input{SystemAmbientBS.tex}}
\label{fig:SystemModel_AmbientBackscatter}%
}\\[5mm]
\subfloat[monostatic backscatter (e.g., RFID)]{\centering
\resizebox{.9\columnwidth}{!}{\input{SystemPassiveRFID.tex}}
\label{fig:SystemModel_RfidBackscatter}%
}
\caption{Circuit descriptions of different classes of BC links. In each case, a load-modulating passive tag (left) transmits to an information receiver (right). 
In (b) the information receiver is also the power source (cf. current $i\Rx$).}
\label{fig:SystemModelCircuit}
\end{figure}

Before studying the information theory of load-modulated BC, we first have to establish an adequate system model. Our approach is based on the circuit models in \Cref{fig:SystemModelCircuit}, which are inspired by \cite{Finkenzeller2015}. They describe the different classes of tag-to-receiver BC links as listed in \cite[Fig.~2]{HuynhCST2018}.
In each case, the left-hand circuit is a tag that modulates information via an adaptive passive load. We employ a symbol time index $n \in \bbZ$ and denote the load impedance $Z\L[n] \in \bbC$. It must fulfill $\Re(Z\L[n]) \geq 0$ at all times because the load is passive \cite[Sec.~4.1]{Pozar2004}.
The tag current phasor $i\Tx[n] \in \bbC$ depends on $Z\L[n]$. The tag antenna impedance is $R\Tx + jX\Tx$, however its reactance $X\Tx$ is canceled by the serial $-X\Tx$ element (resonance). The right-hand circuit is an information receiver that measures a voltage phasor $v[n] \in \bbC$. The tag and receiver circuits are coupled via the mutual impedance $Z\RxTx \in \bbC$, which encapsulates all aspects of the propagation channel.

The circuit \Cref{fig:SystemModel_AmbientBackscatter} describes both ambient and bistatic backscatter links. These paradigms differ only in the assumptions regarding the voltages
$v\Amb\Tx , v\Amb\Rx \in \bbC$
that are induced by an extrinsic electromagnetic field, generated by some source. In ambient backscatter they are random modulated signals from an ambient source, but in the bistatic case they are unmodulated signals from a dedicated source \cite{HuynhCST2018}. 
In either case, $v\Amb\Tx$ is the crucial cause for any electrical activity at the tag while $v\Amb\Rx$ is receive-side interference.

The monostatic case in \Cref{fig:SystemModel_RfidBackscatter} does not assume any extrinsic source. Instead, the information receiver is the system's power source (e.g., an RFID reader) and the crucial tag-side induced voltage $v\Ind\Tx = Z\RxTx\, i\Rx$ is due to the source current $i\Rx$. A prominent example of monostatic BC is inductive RFID, where $X\Tx, X\Rx, Z\RxTx$ are determined by inductances and where $-X\Tx$ is realized by a resonance capacitor.


We assume that $Z\L[n]$ is piecewise constant over time and that it changes instantaneously at the symbol switching instants. We neglect any signal transients which result for $i\Tx$ and $v$. This is a meaningful assumption if the symbol duration is significantly larger than the time constants of the circuits. Our previous work \cite[Appendix~E]{Dumphart2020} showed that transients do not deteriorate the receive processing of load-modulated signals and, when anticipated, can even improve the SNR.

The noise voltage sequence $v\N[n]$ is white Gaussian noise $v\N[n] \iid \calCN(0,\sigma^2)$ with variance $\sigma^2$, a well-established model for thermal noise \cite{Tse2005}. The samples are statistically independent and identically distributed (iid) for different $n$.


A basic circuit analysis yields the tag current expression
\begin{align}
\iTx[n] &= \f{v\Ind\Tx}{R\Tx + Z\L[n]} 
\label{eq:Current} \, .
\end{align}
The receive voltage in the ambient backscatter case is given by
$v[n] = -Z\RxTx\,\iTx[n] + v\Amb\Rx + v\N[n]$.
In the monostatic backscatter case,
$v[n] = -Z\RxTx\,\iTx[n] + (R\Rx + jX\Rx) i\Rx + v\N[n]$.
To unify these different cases within the same system model, we consider a phase rotation
$e^{j\alpha} = \f{(v\Ind\Tx)^*}{|v\Ind\Tx|}$, a specific receive signal compensation,
and other transformations:
\begin{align}
i[n] &:= e^{j\alpha} \, \iTx[n]
= \f{|v\Ind\Tx|}{R\Tx} \f{1}{1 + z[n]}
\label{eq:TxSignalDef} \, , \\
z[n] &:= Z\L[n] \, / \, R\Tx
\label{eq:NormalizedZ} \, , \\
w[n] &:= -e^{j\alpha} \, v\N[n]
\label{eq:NoiseDef} \, , \\
y[n] &:= -e^{j\alpha} \!\left( v[n] - v\big|_{\iTx = 0, v\N = 0}  \right)
\label{eq:ObservationDef} \, .
\end{align}
The noise $w[n]$ maintains the statistics of $v\N[n]$. The unitless $z[n]$ is the normalized load impedance. The transformation from $v$ to $y$ in \Cref{eq:ObservationDef} could be practically realized via interference cancellation, calibration, and channel estimation. For the ambient backscatter case, where $v\Amb\Tx$ and $v\Amb\Rx$ are unknown modulated signals, this delicate aspect is discussed in \Cref{apdx:amb}. Monostatic backscatter systems face the challenge of canceling the strong self-interference $(R\Rx + jX\Rx) i\Rx$, cf. \cite{Finkenzeller2015}.

For either case, the definitions \Cref{eq:TxSignalDef,eq:NormalizedZ,eq:NoiseDef,eq:ObservationDef} yield a complex-valued, discrete-time signal and noise model:
\begin{align}
y[n] &= Z\RxTx \cdot i[n] + w[n]
\label{eq:SignalModel}
\, , \\
w[n] &\iid \calCN(0,\sigma^2)
\, .
\label{eq:NoiseModel}
\end{align}
The observation $y[n] \in \bbC$ is considered without quantization.

\section{Constraint on the Transmit Current}
\label{sec:constraint}
Backscatter tags are passive and thus limited in their capability to establish a desired transmit current $i[n]$.
Formally, this is due to $\Re(z[n]) \geq 0$ in \Cref{eq:TxSignalDef}.
In the following we determine the set of realizable transmit currents, denoted $i[n] \in \Disk$, as prerequisite for the preceding channel capacity analysis.

\begin{figure}[!ht]
\centering
\ \ \ \ \includegraphics[width=.88\columnwidth]{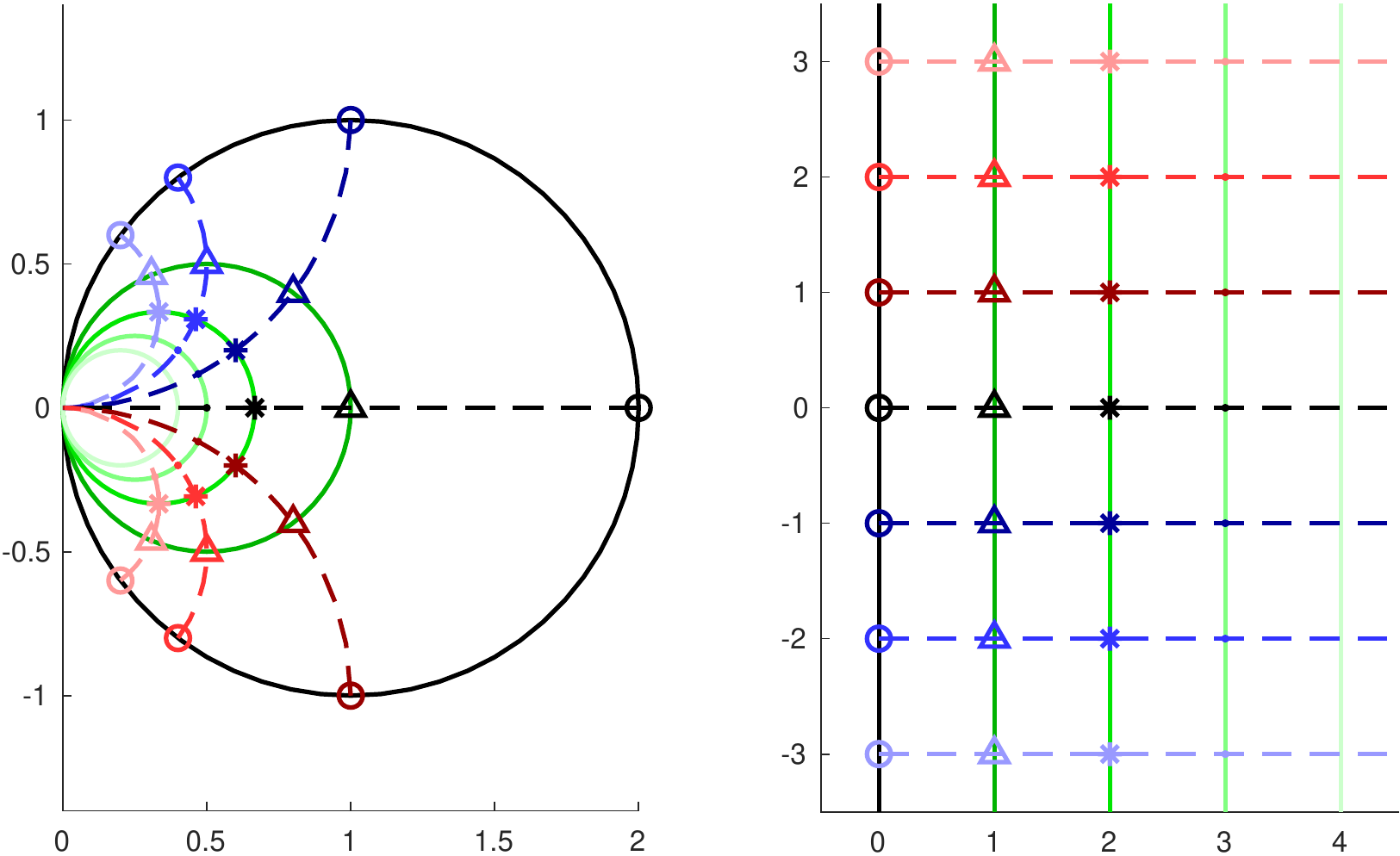}
\put(-230,56){\scriptsize{\rotatebox{90}{$\Im(i) \, / \, \DiskRadius$}}}
\put(-184,-8){\scriptsize{$\Re(i) \, / \, \DiskRadius$}}
\put(-106,62){\scriptsize{\rotatebox{90}{$\Im(z)$}}}
\put(-54,-8){\scriptsize{$\Re(z)$}}
\put(-175,125){\footnotesize{$i = \ChannelFunc(z) = \f{2}{1 + z} \DiskRadius$}}
\caption{Map between normalized load impedance $z$ and transmit current $i$.}
\label{fig:LoadTransformation}
\end{figure}
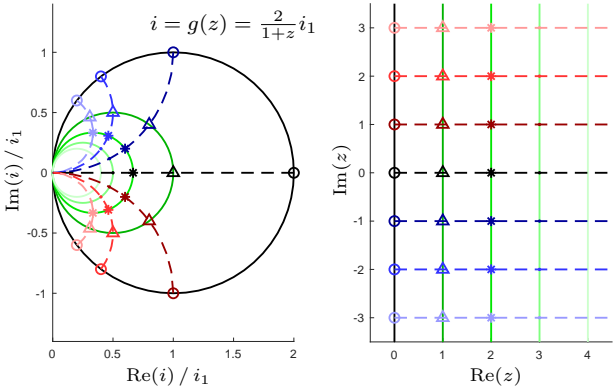

In accordance with typical conventions in communication theory, we henceforth discard time indexation $[n]$ for brevity.
From \Cref{eq:TxSignalDef} we observe that the transmit current $i$ is a non-linear map of the normalized load impedance $z$:
\begin{align}
i &= \ChannelFunc(z) = \f{2\cdot\DiskRadius}{1 + z}
\, , \label{eq:ChannelMap} \\[1mm]
z &= \ChannelFunc^{-1}(i) = \f{2\cdot\DiskRadius}{i} - 1
\, . \label{eq:InverseChannelMap} 
\end{align}
The map $g$ is illustrated in \Cref{fig:LoadTransformation}.
The current quantity $\DiskRadius \in \bbR$ will have the meaning of a radius. It is defined as
\begin{align}
\DiskRadius := \f{|v\Ind\Tx|}{2 R\Tx}
\, . \label{eq:DiskRadius}
\end{align}

The impedance $z$ of any passive load must lie in the right half-plane
$\HalfPlane := \{ z \in \bbC \, | \, \Re(z) \geq 0 \}$. To characterize the set $\Disk = \ChannelFunc(\HalfPlane)$, we rewrite \Cref{eq:ChannelMap} as
$i = \ChannelFunc(z) = \DiskRadius (1 - \f{z-1}{z+1})$ or rather $i = \DiskRadius (1 - \Gamma)$. The reflection coefficient $\Gamma = \f{z-1}{z+1}$ is a bijective map from $z \in \HalfPlane$ to the unit disk $|\Gamma| \leq 1$; it is the M\"obius transformation that also underlies the well-known Smith chart \cite[Eq.~(2.53)]{Pozar2004}. This yields a constraint on the transmit current
\begin{align}
|i - \DiskRadius| \leq \DiskRadius
\label{eq:Constraint}
\end{align}
because $|i- \DiskRadius| = |-\DiskRadius\Gamma| = \DiskRadius |\Gamma| \leq \DiskRadius$.
The set of realizable transmit currents $i \in \Disk$ is thus given by a disk $\Disk \subset \bbC$ with radius $\DiskRadius$ and center $\DiskRadius$:
\begin{align}
\Disk
= \left\{ \ChannelFunc(z) \, \big| \, \Re(z) \geq 0 \right\}
= \left\{ i \in \bbC \ \big| \ |i - \DiskRadius| \leq \DiskRadius \right\} .
\label{eq:Disk}
\end{align}
An analogous observation is found in the  literature, regarding transformed RFID transponder impedance. \cite[Sec.~4.1]{Finkenzeller2015}

\section{Channel Capacity}
\label{sec:capacity}
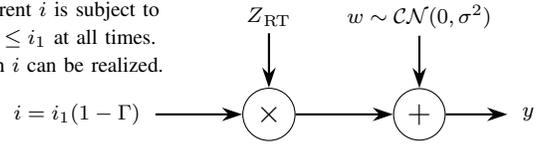
\begin{figure}[t]
\centering
\begin{tikzpicture}[auto,>=latex']
    \node (in) {$i = \DiskRadius (1 - \Gamma)\ $};
    \node [circle, draw, fill=white,right of=in,node distance=25mm] (mult) {\large$\times$};
    \node (constr) [above of=in,node distance=10mm] {$\begin{array}{ll}\text{The current $i$ is subject to}\\[.6mm]\left|i - \DiskRadius\right| \leq \DiskRadius\text{ at all times.}\ \ \ \ \ \ \ \ \ \ \\[.6mm]\text{All such $i$ can be realized.}\end{array}$};
    \node [circle, draw, fill=white] (add) [right of=mult,node distance=20mm] {\large$+$};
    \node (zrt) [above of=mult,node distance=13mm] {$Z\RxTx$};
    \node (w) [above of=add,node distance=13mm] {$w \sim \mathcal{CN}(0,\sigma^2)$};
    \node (out) [right of=add, node distance=14mm]{$\ y$};
    \path[-{Stealth[length=2.7mm,width=1.8mm]},thick] (in) edge node {} (mult);
    \path[-{Stealth[length=2.7mm,width=1.8mm]},thick] (mult) edge node {} (add);
    \path[-{Stealth[length=2.7mm,width=1.8mm]},thick] (zrt) edge node {} (mult);
    \path[-{Stealth[length=2.7mm,width=1.8mm]},thick] (w) edge node {} (add);
    \draw[-{Stealth[length=2.7mm,width=1.8mm]},thick] (add) edge node {} (out) ;
\end{tikzpicture}
\caption{Communication-theoretic description of load modulation in AWGN.}
\label{fig:BlockDiagram}
\end{figure}

We recall the signal model $y = Z\RxTx \cdot i + w$ from \Cref{eq:SignalModel} and the constraint $|i - \DiskRadius| \leq \DiskRadius$ from \Cref{eq:Constraint}. Thereby $Z\RxTx$ is non-random and $w \sim \calCN(0,\sigma^2)$ is additive white Gaussian noise (AWGN). In this simple abstraction, which is visualized in \Cref{fig:BlockDiagram}, the impedance $z$ and the map $\ChannelFunc$ do not occur anymore.

Henceforth, the transmit current $i$ is considered as random variable; the probability density function (PDF) is denoted $f_i$. Its support must lie in the disk, i.e. $\supp(f_{i}) \subseteq \Disk$.

We are interested in the achievable information rates for a given signal-to-noise ratio (SNR). We define the SNR as
\begin{align}
\SNR := \f{|Z\RxTx|^2 \, \DiskRadius^2}{\sigma^2}
\, . \label{eq:SNR}
\end{align}
From an engineering perspective, reliable communication is possible over the channel at any achievable information rate, stated in bit per channel use (bpcu). The largest achievable rate defines the channel capacity $\RateMax$. Suitable error-correcting codes with a very large block length allow for information rates arbitrarily close to $\RateMax$ and with an arbitrarily small block error rate.
Formally, the mutual information $\Rate$ specifies an achievable rate, and the channel capacity $\RateMax$ is the supremum of $\Rate$ over all eligible transmit PDFs $f_i$. \cite{Tse2005}

%

%

A crucial insight is that the disk constraint $|i - \DiskRadius| \leq \DiskRadius$ is equivalent to a peak-power-type constraint $|i - \DiskRadius|^2 \leq \DiskRadius^2$ on the signal $i - \DiskRadius$. The $\DiskRadius$-offset does not affect mutual information. Hence, the capacity problem at hand is equivalent to that of the complex-valued, peak-power constrained AWGN channel. The latter has been solved in \cite{ShamaiTIT1995}.

\subsection{Capacity-Achieving Transmit Current Statistics}
\label{sec:CurrentStats}

\begin{figure}[!ht]
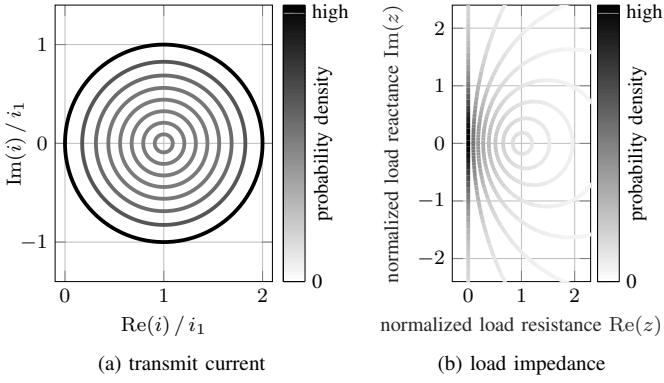

\centering
\subfloat[transmit current]{%
\resizebox{!}{46mm}{\input{CapacityAchievingDistr_sCircles.tex}}
\label{fig:CapAchievingDistr_i}}
\subfloat[load impedance]{
\resizebox{!}{46mm}{\input{CapacityAchievingDistr_zL.tex}}
\label{fig:CapAchievingDistr_z}}
\caption{Capacity-achieving distribution of the relevant complex-valued quantities, plotted for $\SNR = 21\dB$.}
\label{fig:CapAchievingDistr}
\end{figure}

We translate the results in \cite{ShamaiTIT1995} to our formalism. This readily allows for a characterization of the capacity-achieving distribution on the transmit current $i$. It fulfills:
\begin{enumerate}
\item
The PDF support is given by a finite union 
$\supp(f_{i}) = \calC_1 \cup \ldots \cup \calC_K$
of concentric circles $\calC_k$, $k \in \{ 1 , \ldots , K \}$, with radii $\Radius_k$. All circles have their center at $\DiskRadius$.
More formally, the circles are given by
\begin{align}
\calC_k = \left\{\, \DiskRadius + \Radius_k\, e^{j\phi} \, \big| \, \phi \in (-\pi,\pi] \,\right\} .
\label{eq:CircleDefinition}
\end{align}
We assume that the indexing asserts
$0 \leq \Radius_K < \ldots < \DiskRadius$.
\item
The circles are chosen with non-uniform probabilities, denoted as $\CircProb_k$.
\item
The angle $\phi$ has uniform distribution
$\phi \sim \calU(-\pi,\pi)$
for any circle $k$ and for any $\SNR$.
\item
The set of circles always contains the largest possible circle $\calC_1 = \partial\Disk$ (the disk boundary with radius $\DiskRadius$).
\item
The number of circles $K \in \bbN$ increases with $\SNR$.
\item
At low SNR, $K=1$ achieves capacity. This corresponds to a uniform-PSK modulation $i \sim \calU(\calC_1)$.
\end{enumerate}
\Cref{fig:CapAchievingDistr_i} shows a high-SNR example of the capacity-achieving distribution of $i$.

\subsection{Channel Capacity Statement}

For the moment we consider that $K$ circles are given in terms of their radii $\Radius_k$ and probabilities $\CircProb_k$ (their $\SNR$-dependent evolution is covered in the next subsection).
The resulting achievable information rate in bpcu is given by the mutual information expression
\begin{align}
\Rate =&
-\int_0^\infty\! a \cdot \gamma(a) \, \log_2\!\big( \gamma(a) \big)\,da - \log_2(e)
\label{eq:RateFromCircles} \, , \\
\gamma(a) :=&
\sum_{k = 1}^K \,
\CircProb_k \,\exp\!\left( -\f{a^2 + a_k^2}{2} \right)
I_0\!\left( a a_k \right) \, , \\
a_k :=& \f{\Radius_k}{\DiskRadius} \sqrt{2\, \SNR}
\, .
\end{align}
Thereby $I_0$ is the modified Bessel function of the first kind and $e$ is the Euler number.
The integral is evaluated numerically. 
The expression \Cref{eq:RateFromCircles} was obtained by adapting the statements \cite[Eq.~(4),(13),(11),(46)]{ShamaiTIT1995}, which relate to the peak-power constrained quadrature AWGN channel, to our system model. More formal detail is given in \Cref{apdx:ar}.

As stated in \cite{ShamaiTIT1995}, the channel capacity $\RateMax$ is obtained by maximizing $\Rate$ with respect to the free circle parameters:
\begin{align}
\RateMax = &\max_{\Radius_2\, , \ldots , \,\Radius_K, \,\CircProb_1\, , \ldots, \,\CircProb_K} \Rate
\label{eq:Capacity} \\
&\mathrm{subject\ to}\
0 \leq \Radius_K < \ldots < \Radius_2 < \Radius_1 \, , \nonumber\\
&\hphantom{\mathrm{subject\ to}}\
\CircProb_1 , \ldots , \CircProb_K \in [0,1] \, , \nonumber\\
&\hphantom{\mathrm{subject\ to}}\
\CircProb_1 + \ldots + \CircProb_K = 1 \, . \nonumber
\end{align}
In all following evaluations, this problem is tackled with an interior-point algorithm for constrained nonlinear optimization \cite{InteriorPointAlgorithm}, with sensible choices for the initial values.



\subsection{Optimal Number of Circles}
\label{sec:K}

We have yet to address the optimal number of circles $K \in \bbN$ for a given $\SNR$. The intervals where $K = 1,2,3$ are optimal are stated in \cite[Tab.~1]{ShamaiTIT1995} (please note that their $\SNR$ thresholds are $3\dB$ larger because they assumed an AWGN variance of $2$).
A very important fact is the optimality of $K = 1$ for $\SNR < 3.011$ or rather $\SNR < 4.8\dB$ (this threshold was originally determined by solving \cite[Eq.~(45)]{ShamaiTIT1995} numerically).
Beyond that, the optimal $K$ increases with $\SNR$ \cite{ShamaiTIT1995}.

\begin{figure}[t]
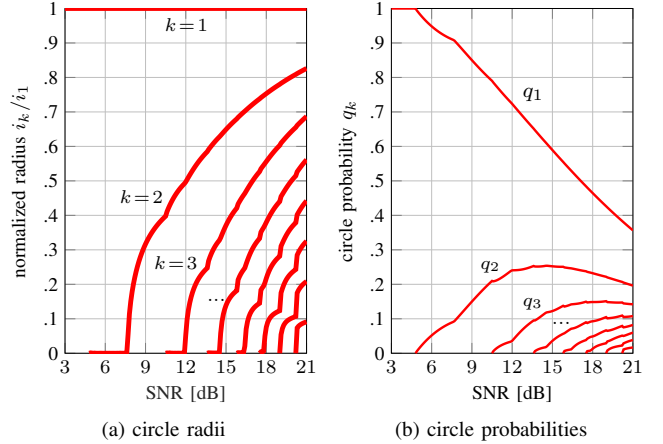

\centering
\subfloat[circle radii]{%
\resizebox{!}{55mm}{\input{CapacityAchievingCircles_Rad}}%
\put(-60,143){\scriptsize{$k \!=\! 1$}}%
\put(-77,78){\scriptsize{$k \!=\! 2$}}%
\put(-64,53){\scriptsize{$k \!=\! 3$}}%
\put(-44,41){\scriptsize{$...$}}%
\label{fig:Radii_evolution}%
}
\
\subfloat[circle probabilities]{%
\resizebox{!}{55mm}{\input{CapacityAchievingCircles_Prob}}%
\put(-48,118){\footnotesize{$\CircProb_1$}}%
\put(-64,53){\scriptsize{$\CircProb_2$}}%
\put(-48,39){\scriptsize{$\CircProb_3$}}%
\put(-37,32){\scriptsize{$...$}}%
\label{fig:PMF_evolution}%
}
\caption{Evolution of the radii and probabilities of the circles that describe the capacity-achieving transmit distribution.}
\label{fig:Radii_PMF_evolution}
\end{figure}

\Cref{fig:Radii_PMF_evolution} shows how new circles emerge with increasing $\SNR$. For this numerical evaluation we iterated through a fine grid of increasing $\SNR$ values. For each $\SNR$ value, all $\Radius_k , \CircProb_k$ were optimized according to \Cref{eq:Capacity}, whereby their preceding values were used as initial values. We added a new smallest circle whenever this addition caused an appreciable rate increase. The associated numerical thresholds have a noticeable effect in the high-SNR regime, because there, parameter fine tuning of the innermost circles only causes rate changes near the floating point accuracy. In detail, we required that a new $K$-th circle must have probability $\CircProb_K \geq 0.003 \cdot \CircProb_{K-1}$ and must yield a rate increase larger than $100$ times the floating-point relative accuracy of Matlab ($2^{-52}$). 


\begin{figure}[t]
\centering
\resizebox{.96\columnwidth}{!}{\begin{tikzpicture}

\begin{axis}[%
width=90mm,
height=45mm,
at={(0,0)},
scale only axis,
xmin=-10,
xmax=24,
xtick={ -9, -6, -3, 0, 3, 6, 9, 12, 15, 18, 21, 24},
xlabel style={font=\color{white!15!black}},
xlabel={$\SNR$ $[\mathrm{dB}]$},
ymode=log,
ymin=0.1,
ymax=10,
yminorticks=true,
ylabel style={font=\color{white!15!black}},
ylabel={channel capacity [bpcu]},
axis background/.style={fill=white},
xmajorgrids,
ymajorgrids,
yminorgrids,
grid style={opacity=0.13},
minor grid style={opacity=0.3},
label style={font=\normalsize},
legend style={at={(1,0.085)}, anchor=south east, legend cell align=left, align=left, draw=white!15!black}
]
\addplot [color=blue, dashed, line width=0.5pt]
  table[row sep=crcr]{%
-10	0.137503523749935\\
-9.5	0.153418636061023\\
-9	0.171069138514398\\
-8.5	0.190619606352057\\
-8	0.212244743260395\\
-7.5	0.236128803353125\\
-7	0.262464707140878\\
-6.5	0.291452814285687\\
-6	0.323299322693842\\
-5.5	0.358214274110159\\
-5	0.396409161163114\\
-4.5	0.438094149757986\\
-4	0.48347495334568\\
-3.5	0.532749420867551\\
-3	0.586103926445348\\
-2.5	0.643709673944082\\
-2	0.705719050773513\\
-1.5	0.772262179969205\\
-1	0.843443825203652\\
-0.5	0.91934079807483\\
0	1\\
0.5	1.0854372028192\\
1	1.17563663469239\\
1.5	1.27055139420231\\
2	1.37010466975099\\
2.5	1.47419169766592\\
3	1.58268235491156\\
3.5	1.69542425407813\\
4	1.81224619130063\\
4.5	1.9329617924573\\
5	2.0573732086068\\
5.5	2.18527472629821\\
6	2.31645617962626\\
6.5	2.45070607596247\\
7	2.58781437356203\\
7.5	2.72757487451865\\
8	2.86978721917029\\
8.5	3.01425848700632\\
9	3.16080442391302\\
9.5	3.30925032620402\\
10	3.4594316186373\\
10.5	3.61119416705619\\
11	3.76439436704286\\
11.5	3.91889904869198\\
12	4.07458523490543\\
12.5	4.23133978700232\\
13	4.38905896736305\\
13.5	4.54764794461157\\
14	4.70702026272884\\
14.5	4.86709729164214\\
15	5.02780767335052\\
15.5	5.18908677457204\\
16	5.3508761542486\\
16.5	5.51312305200527\\
17	5.67577990180488\\
17.5	5.83880387352813\\
18	6.00215644400198\\
18.5	6.16580299805362\\
19	6.32971245944191\\
19.5	6.49385695097752\\
20	6.65821148275179\\
20.5	6.82275366712554\\
21	6.98746345895592\\
21.5	7.1523229194428\\
22	7.31731600193655\\
22.5	7.48242835805457\\
23	7.64764716249041\\
23.5	7.81296095495953\\
24	7.97835949780125\\
};
\addlegendentry{$\text{upper bound log}_\text{2}\text{(1 + SNR)}$}

\addplot [color=red, line width=1.3pt]
  table[row sep=crcr]{%
-10	0.137490906666417\\
-9.5	0.153399403990599\\
-9	0.171039935249776\\
-8.5	0.190575444753509\\
-8	0.212178257195243\\
-7.5	0.236029182897772\\
-7	0.262316197838123\\
-6.5	0.291232624383457\\
-6	0.322974738129779\\
-5.5	0.357738730230072\\
-5	0.395716964847275\\
-4.5	0.43709348985187\\
-4	0.48203878775833\\
-3.5	0.530703795313945\\
-3	0.583213275981814\\
-2.5	0.639658700930992\\
-2	0.700090880847482\\
-1.5	0.764512690550033\\
-1	0.832872335579065\\
-0.5	0.905057715083341\\
0	0.980892523998895\\
0.5	1.06013479000486\\
1	1.14247853280152\\
1.5	1.22755913836571\\
2	1.31496283468871\\
2.5	1.404240323024\\
3	1.49492416330459\\
3.5	1.58654896613884\\
4	1.67867287507681\\
4.5	1.77089833749213\\
5	1.86336708293004\\
5.5	1.95951129500193\\
6	2.05944392499972\\
6.5	2.16227922545915\\
7	2.26702644878706\\
7.5	2.37263814538457\\
8	2.47879831992885\\
8.5	2.58745529907544\\
9	2.69871645260338\\
9.5	2.81240690699709\\
10	2.92808134803176\\
10.5	3.0452964353854\\
11	3.16430965375948\\
11.5	3.2854181526281\\
12	3.40818887394725\\
12.5	3.53259192319178\\
13	3.65883219897967\\
13.5	3.78674553931811\\
14	3.91623260483693\\
14.5	4.04733161815581\\
15	4.17992003073123\\
15.5	4.31401399928841\\
16	4.44948247348643\\
16.5	4.5860705935919\\
17	4.72339744708351\\
17.5	4.86097776801742\\
18	4.99826548372716\\
18.5	5.13471323627153\\
19	5.26983776127836\\
19.5	5.40327367267533\\
20	5.53479284393184\\
20.5	5.66428043683202\\
21	5.79169176380879\\
21.5	5.91702209959332\\
22	6.04029376930273\\
22.5	6.16154963982862\\
23	6.28084773802992\\
23.5	6.39825660758257\\
24	6.51385156163464\\
};
\addlegendentry{channel capacity, general passive load}

\addplot [color=black, dashdotted, line width=0.5pt]
  table[row sep=crcr]{%
-10	0.137490906666417\\
-9.5	0.153399403990599\\
-9	0.171039935249776\\
-8.5	0.190575444753509\\
-8	0.212178257195243\\
-7.5	0.236029182897772\\
-7	0.262316197838123\\
-6.5	0.291232624383457\\
-6	0.322974738129779\\
-5.5	0.357738730230072\\
-5	0.395716964847275\\
-4.5	0.43709348985187\\
-4	0.48203878775833\\
-3.5	0.530703795313945\\
-3	0.583213275981814\\
-2.5	0.639658700930992\\
-2	0.700090880847482\\
-1.5	0.764512690550033\\
-1	0.832872335579065\\
-0.5	0.905057715083341\\
0	0.980892523998895\\
0.5	1.06013479000486\\
1	1.14247853280152\\
1.5	1.22755913836571\\
2	1.31496283468871\\
2.5	1.404240323024\\
3	1.49492416330459\\
3.5	1.58654896613884\\
4	1.67867287507681\\
4.5	1.77089833749213\\
5	1.86288989247773\\
5.5	1.95438678032001\\
6	2.04520868673105\\
6.5	2.13525386861277\\
7	2.22449012886201\\
7.5	2.3129403489088\\
8	2.40066521289591\\
8.5	2.48774607413584\\
9	2.57427050396812\\
9.5	2.66032206006254\\
10	2.74597457231891\\
10.5	2.83129019850657\\
11	2.91631995217247\\
11.5	3.00110540225082\\
12	3.08568060483823\\
12.5	3.17007378252874\\
13	3.25430861502594\\
13.5	3.3384051836574\\
14	3.42238066188835\\
14.5	3.50624983219313\\
15	3.59002548434695\\
15.5	3.67371873004793\\
16	3.75733925624968\\
16.5	3.84089553217357\\
17	3.92439498047657\\
17.5	4.00784412016336\\
18	4.09124868688507\\
18.5	4.17461373491025\\
19	4.25794372408082\\
19.5	4.34124259434809\\
20	4.42451382995459\\
20.5	4.50776051491685\\
21	4.59098538115687\\
21.5	4.67419085038305\\
22	4.7573790706311\\
22.5	4.84055194822161\\
23	4.92371117576761\\
23.5	5.00685825676705\\
24	5.08999452723369\\
};
\addlegendentry{channel capacity, purely reactive load\\(equivalent: uniform-PSK modulation)}
\end{axis}

\end{tikzpicture}%
}%
\caption{Channel capacity in bit per channel use (bpcu) plotted versus $\SNR$.} 
\label{fig:Rates_Capacity}
\end{figure}
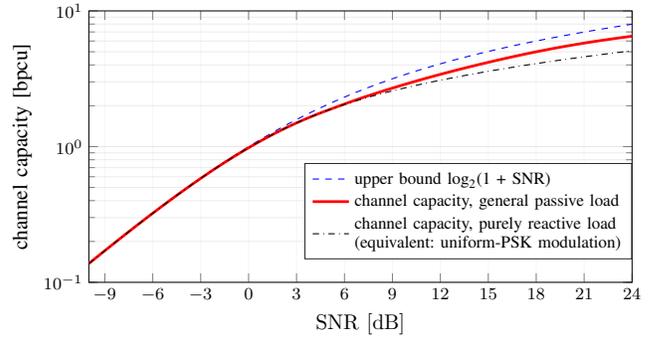

\Cref{fig:Rates_Capacity} plots $\Rate$ versus $\SNR$ for different assumptions:
\begin{itemize}
\item \textbf{The capacity-achieving $K$ and $\Radius_k$, $\CircProb_k$ are used:} At high SNR, the additional circles inside the disk ensure that the transmit signal space is utilized thoroughly. 
This is achieved by purposefully raising the load resistance $\Re(z) \geq 0$.
\item \textbf{Only $K = 1$ is used:} This uniform-PSK $i \sim \calU(\calC_1)$ at the disk boundary is associated with a purely reactive load ($z = jx$), which is an interesting feature from a circuit perspective. The resulting information rate also constitutes the channel capacity of reactive load modulation. At high SNR, it lacks behind general passive load modulation, because the disk interior is not utilized.
\end{itemize}
For $\SNR < 4.8\dB$ the two cases coincide precisely, because then $K=1$ achieves capacity.

\subsection{Capacity-Achieving Load Impedance Statistics}
\label{sec:loadstats}

Herein we characterize the capacity-achieving distribution of the load impedance $z$, which follows from $z = \ChannelFunc^{-1}(i)$ together with the statistics of $i$ described in \Cref{sec:CurrentStats}.

The circle $\calC_1$ maps to the imaginary axis $j\bbR = \ChannelFunc^{-1}(\calC_1)$, i.e. to the set of purely reactive loads.
In detail, $i = \DiskRadius + \DiskRadius\, e^{j\phi}$ maps to $z = jx$ with
$x = -\tan(\phi / 2)$ and $\phi \sim \calU(-\pi,\pi)$. We find that $x$ has standard Cauchy distribution; the PDF is
\begin{align}
f_x(x) &= \f{1}{\pi(1 + x^2)} \, , &
x &\in \bbR \, . &
\label{eq:Lorentz}
\end{align}
This is a complete description of the capacity-achieving load statistics in the low-SNR case (where $K = 1$ is optimal) or when a purely reactive load is enforced for technical reasons. The specific probability density evolution from \Cref{eq:Lorentz} can also seen along the imaginary axis in \Cref{fig:CapAchievingDistr_z}.


The inner circles $\calC_k$ with $k \in \{2,\ldots,K\}$ map to circles $z \in \ChannelFunc^{-1}(\calC_k)$ with centers
$\f{\Radius_1^2 + \Radius_k^2}{\Radius_1^2 - \Radius_k^2} \in \bbR$
and radii
$\f{2 \Radius_1 \Radius_k}{\Radius_1^2 - \Radius_k^2}$.
In detail, a current phasor $i = \DiskRadius + \Radius_k\, e^{j\phi} \in \calC_k$ maps to
\begin{align}
z &= g^{-1}\left( i \right)
= \f{\Radius_1^2 + \Radius_k^2}{\Radius_1^2 - \Radius_k^2} + \f{2 \Radius_1 \Radius_k}{\Radius_1^2 - \Radius_k^2} \, e^{j(\theta + \pi)}
, \label{eq:z_from_s} \\
\theta &= 2 \arctan\left( \f{\sin\phi}{\Radius_k  / \Radius_1 + \cos\phi} \right) - \phi
\, . \label{eq:phi_from_phi}
\end{align}
If $\Radius_k / \Radius_1 \ll 1$ then the approximate linearity $\theta \approx \phi$ holds.
If $\Radius_k / \Radius_1 \approx 1$ then $\theta$ is pushed towards zero.
Both properties can be observed in the high-SNR example in \Cref{fig:CapAchievingDistr_z}.
The distribution of $\theta|k$ is determined by \Cref{eq:phi_from_phi} and
$\phi \sim \calU(-\pi,\pi)$.%




To draw samples $Z\L$ from the capacity-achieving distribution, the following simple procedure suffices. Choose a circle $k$ according to the probabilities $\CircProb_k$ and draw a sample of the angle $\phi \sim \calU(-\pi,\pi)$. Compute $i = \DiskRadius + \Radius_k\, e^{j\phi}$, \mbox{$z = \ChannelFunc^{-1}(i)$}, and finally $Z\L = R\Tx \cdot z$. This way, a capacity-achieving codebook of load impedances $Z\L$ can be obtained.

\subsection{Upper Bound on the Capacity}

The effective constraint $|i - \DiskRadius|^2 \leq \DiskRadius^2$ of the peak-power type is obviously stricter than a constraint $\EV{|i - \DiskRadius|^2} \leq \DiskRadius^2$ of the average-power type. This inflicts the upper bound \cite{ShamaiTIT1995}
\begin{align}
\RateMax < \log_2(1 + \SNR).
\label{eq:UpperBound}
\end{align}
\Cref{fig:Rates_Capacity} shows that the bound is practically tight at low $\SNR$. Formally however, equality is ruled out by the following argument. By \cite[Appendix~B.4]{Tse2005}, equality would require a Gaussian $i \sim \mathcal{CN}(\DiskRadius, \DiskRadius^2)$ whose PDF support $\bbC \nsubseteq \Disk$ however violates the disk constraint \Cref{eq:Constraint}.


\section{Near-Capacity Rates with Finite Alphabets}
\label{sec:practical}
Most every practical digital modulation uses a finite symbol alphabet instead of a continuous transmit distribution.
In our formalism this means that $i$ is chosen from an alphabet $i \in \SymbolAlphabet$, $\SymbolAlphabet = \{ s_1 , \ldots , s_M \} \subset \Disk$,
associated with $z \in \SymbolAlphabetZL$ from a load impedance alphabet $\SymbolAlphabetZL = \ChannelFunc^{-1}(\SymbolAlphabet) = \{ z_1 , \ldots , z_M \} \subset \HalfPlane$.
This caps the achievable rate at $\log_2(M)$ bpcu.
The Euclidean symbol distance $|s_m - s_n| \leq 2\DiskRadius$ is capped by the disk diameter.
This maximum is attained by a binary alphabet 
$\SymbolAlphabet = \{ 0, 2\DiskRadius \}$,
$\SymbolAlphabetZL = \{ \infty , 0\}$
but also by
$\SymbolAlphabet = \{ \DiskRadius(1+j) , \DiskRadius(1-j) \}$,
$\SymbolAlphabetZL = \{ -j , +j \}$
or any rotation of such $\SymbolAlphabet$ about $\DiskRadius$.

\begin{figure}[!ht]
\centering
\vspace{-5mm}
\subfloat[symbol alphabet, TX current]{%
\resizebox{!}{60mm}{\input{Constellation_i.tex}\vspace{-1mm}}
\label{fig:Constellation_i}}\!\!\!\!
\subfloat[load impedance alphabet]{
\resizebox{!}{60mm}{\input{Constellation_z.tex}\vspace{-1mm}}
\label{fig:Constellation_z}}\\[2mm]
\subfloat[resulting information rate]{
\includegraphics[width=.85\columnwidth]{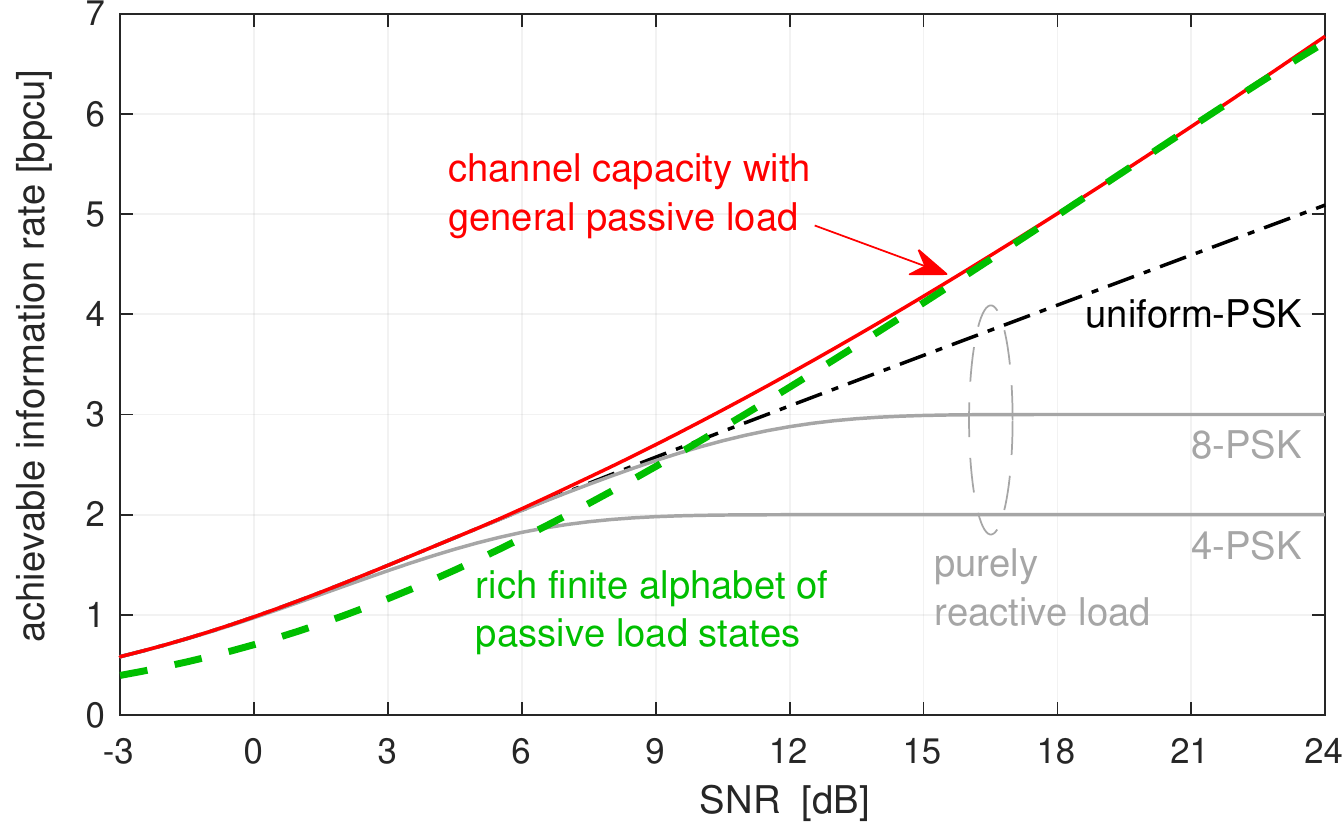}
\label{fig:Constellation_rate}}
\caption{Proposal for a rich 8-bit symbol alphabet that allows for near-capacity information rates at high $\SNR$.}
\label{fig:Constellation}
\end{figure}

$M$-ary phase shift keying ($M$-PSK) is a more capable alphabet. It uses $M = 2^\ell$ symbols 
at $s_m = \DiskRadius(1+\exp(j2\pi\f{m - 1/2}{M}))$.
As mentioned earlier, this modulation is realized with a purely reactive load circuit.
\Cref{fig:Constellation_rate} shows the achievable rate of various PSK schemes. 
The underlying numerical rate computation is described in \Cref{apdx:ar}.
At low SNR, $M$-PSK with $M \geq 4$ yields data rates very close to channel capacity while uniform-PSK ($M = \infty$) even achieves capacity.

The high-SNR gap between PSK and channel capacity confirms that purposefully adding load resistance is crucial for achieving very high data rates. We are interested in a rich symbol alphabet that remedies this gap. Inspired by \Cref{fig:CapAchievingDistr}, the capacity-achieving distribution at $\SNR = 21\dB$, we propose the symbol alphabet in \Cref{fig:Constellation_i}. It uses $M = 2^8 = 256$ and a heuristic construction that ensures large pairwise symbol distances. The symbol probabilities are set such that the outer circle is chosen with $\CircProb_1 = 0.36$ (the high-SNR value from \Cref{fig:PMF_evolution}). The associated rate graph (green dashed) in \Cref{fig:Constellation_rate} indeed demonstrates high-SNR rates very close to channel capacity. The low-SNR gap could be closed by adapting the symbol probabilities $q_m$ to the SNR (like in adaptive modulation), which is omitted for brevity.

The constellation plot \Cref{fig:Constellation_i} highlights certain symbols that are unrealizable when the load is subject to certain value-range constraints. This particular evaluation assumes an inductive RFID tag whose coil antenna ($Z\Tx = R\Tx + j\omega L\Tx$) is loaded with an impedance $R+\f{1}{j\omega C}$ with adaptive $R,C \in \bbR_+$, from the value range
$(1-\Delta)C_\text{res} \leq C \leq (1+\Delta)C_\text{res}$
about the resonance value
$C_\text{res} = 1/(\omega^2 L\Tx)$.
It can be shown that this is equivalently described by our \Cref{sec:model} model with
$z = r + jx$
and
$x = x\Tx (1 - \f{C_\text{res}}{C})$
subject to
$\f{-\Delta}{1-\Delta} x\Tx \leq x \leq \f{\Delta}{1+\Delta} x\Tx$. Thereby $x\Tx = X\Tx / R\Tx = \omega L\Tx / R\Tx$ is the coil Q-factor.
In \Cref{fig:Constellation_i} we assume $\Delta = 0.5$ and $x\Tx = 15$, which yields 9 out of 256 unrealizable symbols. The resultant loss of achievable rate turns out to be negligible at the considered SNR range (the graph is not shown in \Cref{fig:Constellation_rate} because visually it coincides with the dashed green graph). We conclude that mild value-range constraints do not prohibit near-capacity data rates.

\section{Summary}
\label{sec:summary}
For the first time this paper stated the channel capacity of load modulation with a freely adaptable passive impedance. The obtained insights on the capacity-achieving transmit distribution and how to approximate it with finite symbol alphabets have important implications for practical high-data-rate backscatter communication systems. This applies even to the ambient backscatter case, under certain identified conditions.

\begin{appendices}
\crefalias{section}{appendix}
	
    \section{Conditions on Modulated Ambient Signals}
    \label{apdx:amb}
    The ambient backscatter case requires special care because the voltages $v\Tx\Amb$ and $v\Rx\Amb$ may exhibit fast time-variations from modulation.
In that regard, we require the following conditions:
\mbox{(i) The} relevant propagation channels are either subject to block fading or no fading at all.
(ii) There is only a single ambient source and it uses digital modulation.
(iii) The channel from ambient source to receiver is much stronger than the backscatter channel, i.e. $\EV{|Z\RxTx i\Tx|^2} \ll \EV{|v\Rx\Amb|^2}$.
(iv) The modulated signal $v\Rx\Amb$ can be decoded correctly.
(v) There is no interference from other backscatter tags.

We identify the following different cases for which the channel capacity result \Cref{eq:Capacity} applies to ABC in some fashion:

\textbf{1.) The ambient source has much faster symbol rate than the load modulation:} Let $L \gg 1$ denote the ratio of symbol rates and assume $L \in \bbN$. We consider the fast symbol rate with time index $\ell$. Let $s[\ell] := \f{-Z\RxTx}{R\Tx(1 + z[\ell])} v\Tx\Amb[\ell] + v\N[\ell]$, which is $v$ after compensation of the decoded $v\Rx\Amb[\ell]$. The modulated $v\Tx\Amb[\ell]$ is i.i.d. random and $v\N[\ell] \iid \calN(0,\sigma^2 L)$ while $z[\ell]$ is constant over length-$L$ blocks. For a specific block 
we collect the various signals in the vectors ${\bf s}, {\bf v}\Tx\Amb, {\bf v}\N \in \bbC^L$ to write
${\bf s}
= \f{-Z\RxTx}{R\Tx(1 + z)} {\bf v}\Tx\Amb + {\bf v}\N$.
We consider maximum-ratio combining
$\tilde{y} = - {\bf u}\H {\bf s} / \sqrt{L}$ at the receiver, whereby
${\bf u} := {\bf v}\Tx\Amb / \|{\bf v}\Tx\Amb\|$.
This results in the relation $\tilde{y} = \f{Z\RxTx}{R\Tx(1 + z)} \|{\bf v}\Tx\Amb\| / \sqrt{L} + \omega$ with $\omega \sim \calN(0,\sigma^2)$. This relation is equivalent to the signal model \Cref{eq:SignalModel} with the exception that $|v\Tx\Amb|$ is replaced by $\|{\bf v}\Tx\Amb\| / \sqrt{L}$. The latter approaches the RMS value of $v\Tx\Amb$ for large $L$. Therefore the system behaves as if $v\Tx\Amb$ was constant.

\textbf{2.) The ambient source has much slower symbol rate than the load modulation:} The effect on the backscatter system is the same as if $v\Tx\Amb$ was unmodulated but subject to block fading. By coding across many such blocks, the information rate $\EVVar{v\Tx\Amb}{\RateMax}$ can be achieved \cite[Sec.~5.4.5]{Tse2005}. Thereby $\RateMax$ is the complicated expression from \Cref{eq:Capacity}.

\textbf{3.) The ambient source is PSK modulated:} PSK has a constant envelope, so $|v\Tx\Amb|$ and $|v\Rx\Amb|$ are constant for the duration of a fading block. The phase shifts in $v$ due to $v\Tx\Amb$ can be compensated with the knowledge from the decoded $v\Rx\Amb$. Then the system behaves as if $v\Tx\Amb$ was constant.

    \section{Mutual Information Details}
    \label{apdx:ar}
    The signal model $y = Z\RxTx \cdot i + w$ from \Cref{eq:SignalModel} exhibits AWGN $w \iid \mathcal{CN}(0,\sigma^2)$ and non-random $Z\RxTx$. The achievable rate is given by the mutual information $\Rate$ between output $y$ and input $i$. For a complex-valued AWGN channel \cite[(B.47)]{Tse2005}
\begin{align}
& \Rate = h(y) - \log_2(\pi e \sigma^2)
\, , \label{eq:MutualInfo} \\
& h(y) = -\int_\bbC f_y(y) \, \log_2\big( f_y(y) \big) \,dy
\label{eq:DifferentialEntropy}
\end{align}
where $h(y)$ is the differential entropy of $y$.
The PDF $f_y = f_v * f_w$ is the convolution of the PDF $f_v$ of the receive-side signal $v := Z\RxTx \cdot i$ and the PDF
$f_w(w) = \f{1}{\pi \sigma^2} \exp(-\f{|w|^2}{\sigma^2})$.

With the transmit statistics from \Cref{sec:CurrentStats}, $f_i$ and $f_v$ are supported on concentric circles. Then $f_y$ is characterized by noise-convoluted concentric circles. Conditioned on circle $k$, $|y|$ has Rice distribution and $\arg(y)$ uniform distribution. These facts can be used in \Cref{eq:DifferentialEntropy} to derive the result \Cref{eq:RateFromCircles} for $\Rate$ after lengthy calculation, as detailed in \cite{ShamaiTIT1995}.

When $i$ is instead chosen from a finite symbol alphabet $\{ s_1 , \ldots , s_{M} \} \subseteq \Disk$ with probabilities $\SymbProb_m$, then
$f_v(v) = \sum_{m=1}^M \SymbProb_m \delta(v - Z\RxTx s_m)$.
The convolution with $f_w$ yields
$f_y(y) = \sum_{m=1}^M \f{\SymbProb_m}{\pi \sigma^2} \exp(-\f{1}{\sigma^2} |y - Z\RxTx s_m|^2 )$.
This expression allows to calculate the differential entropy $h(y)$ with numerical integration to then evaluate the mutual information \Cref{eq:MutualInfo}.

\end{appendices}

\bibliographystyle{IEEEtran}
\bibliography{ref}

\pdfinfo{
	   /Author(Gregor Dumphart, Johannes Sager, Armin Wittneben)
	   /Title(\OurPaperTitle{})
	   /Subject(\OurKeywords{})
	   /Keywords(\OurKeywords{})
}

\end{document}